\documentclass{IEEEtran4PSCC}
\usepackage{cite}
\ifCLASSINFOpdf
\usepackage[pdftex]{graphicx}
\DeclareGraphicsExtensions{.pdf,.jpeg,.png}
\else
\fi
\usepackage[cmex10]{amsmath}
\usepackage{url}
\makeatletter
\let\old@ps@headings\ps@headings
\let\old@ps@IEEEtitlepagestyle\ps@IEEEtitlepagestyle
\def\psccfooter#1{%
    \def\ps@headings{%
        \old@ps@headings%
        \def\@oddfoot{\strut\hfill#1\hfill\strut}%
        \def\@evenfoot{\strut\hfill#1\hfill\strut}%
    }%
    \def\ps@IEEEtitlepagestyle{%
        \old@ps@IEEEtitlepagestyle%
        \def\@oddfoot{\strut\hfill#1\hfill\strut}%
        \def\@evenfoot{\strut\hfill#1\hfill\strut}%
    }%
    \ps@headings%
}
\makeatother

\psccfooter{%
        \parbox{\textwidth}{\hrulefill \\ \small{21st Power Systems Computation Conference} \hfill \begin{minipage}{0.2\textwidth}\centering \vspace*{4pt} \includegraphics[scale=0.06]{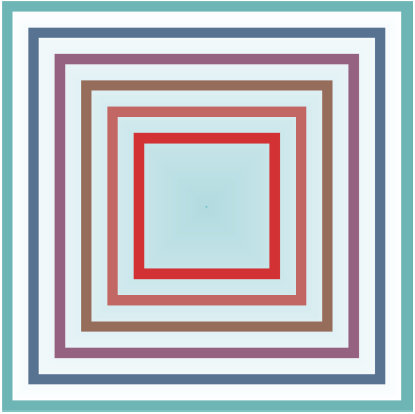}\\\small{PSCC 2020} \end{minipage} \hfill \small{Porto, Portugal --- June 29 -- July 3, 2020}}%
}

\usepackage[nolist,printonlyused]{acronym}
\usepackage{dirtree}
\usepackage{soul}
\usepackage{color}

\begin{document}
\title{Computational Experiment Design for Operations Model Simulation}

%% To specify the authors when (number of affiliations <= 2)
 \author{\IEEEauthorblockN{Jose Daniel Lara\IEEEauthorrefmark{1} \IEEEauthorrefmark{3},
Jonathan T. Lee\IEEEauthorrefmark{1},
Duncan Callaway\IEEEauthorrefmark{1},
Bri-Mathias Hodge\IEEEauthorrefmark{2} \IEEEauthorrefmark{3}}
\IEEEauthorblockA{\IEEEauthorrefmark{1} Energy and Resources Group, University of California Berkeley}
\IEEEauthorblockA{\IEEEauthorrefmark{2} Electrical, Computer \& Energy Engineering and Renewable and Sustainable Energy Institute, University of Colorado Boulder}
\IEEEauthorblockA{\IEEEauthorrefmark{3} National Renewable Energy Laboratory (NREL)}
}

%% To specify the authors when (number of affiliations > 2)
% \author{\IEEEauthorblockN{Author n.1\IEEEauthorrefmark{1},
% Author n.2\IEEEauthorrefmark{2},
% Author n.3\IEEEauthorrefmark{3},
% Author n.4\IEEEauthorrefmark{3} and
% Author n.5\IEEEauthorrefmark{4}}
% \IEEEauthorblockA{\IEEEauthorrefmark{1} Department Name of Organization A\\
% Name of the organization A,
% Address A\\ Emails if wanted}
% \IEEEauthorblockA{\IEEEauthorrefmark{2} Department Name of Organization B\\
% Name of the organization B,
% Address B\\ Emails if wanted}
% \IEEEauthorblockA{\IEEEauthorrefmark{3} Department Name of Organization C\\
% Name of the organization C,
% Address C\\ Emails if wanted}
% \IEEEauthorblockA{\IEEEauthorrefmark{4}Department Name of Organization D\\
% Name of the organization D,
% Address D\\ Emails if wanted}
% }

\maketitle

\begin{abstract}
% Advancing research and changing power systems practices to provide more
Computer simulations that demonstrate the value of novel approaches are crucial to developing more flexible and robust power systems operations with high penetrations of renewable energy at multiple geographic and temporal scales. However, optimization-based simulations that depend on forecast data often face challenges in evaluating performance, reproducing results, and testing under realistic simulation scenarios. In this paper, we develop scientific computing best-practices for the validation and reproduction of power systems operational models. We then employ two case studies to demonstrate the proposed validation and reproduction framework.
\end{abstract}


% Use this to place sponsorships
\thanksto{This work was authored [in part] by the National Renewable Energy Laboratory (NREL), operated by Alliance for Sustainable Energy, LLC, for the U.S. Department of Energy (DOE) under Contract No. DE-AC36-08GO28308. This work was supported by the Laboratory Directed Research and Development (LDRD) Program at NREL. The views expressed in the article do not necessarily represent the views of the DOE or the U.S. Government.}

\begin{acronym}[PSCC]
\acro{AML}{Algebraic Modeling Language}
\acro{CCO}{Chance Constrained Optimization}
\acro{CDF}{Cumulative Probability Distribution Function}
\acro{CHP}{Combined Heat and Power}
\acro{DER}{Distributed Energy Resource}
\acro{DG}{Distributed Generator}
\acro{DSM}{Demand Side Management}
\acro{ED}{Economic Dispatch}
\acro{EMS}{Energy Management System}
\acro{ESS}{Energy Storage System}
\acro{LP}{Linear Programming}
\acro{MC}{Monte Carlo}
\acro{MILP}{Mixed-Integer Linear Programming}
\acro{MP}{Mathematical Program}
\acro{NLP}{Nonlinear Programming}
\acro{OPF}{Optimal Power Flow}
\acro{PEC}{Power Electronics Converter}
\acro{PDF}{Probability Density Function}
\acro{PV}{Photovoltaic}
\acro{PF}{Power Flow}
\acro{RE}{Renewable Energy}
\acro{RHC}{Receding Horizon Control}
\acro{RO}{Robust Optimization}
\acro{RUC}{Robust Unit Commitment}
\acro{SC}{Scientific Computing}
\acro{SO}{Stochastic Optimization}
\acro{SCIG}{Squirrel-cage Induction Generator}
\acro{SoC}{State-of-Charge}
\acro{SP}{Stochastic Programming}
\acro{SUC}{Stochastic Unit Commitment}
\acro{UC}{Unit Commitment}
\acro{VSC}{Voltage Source Converter}
\acro{WT}{Wind Turbine}
\end{acronym}

\section{Introduction}

The study of power systems operations has evolved with the changing needs of the grid and the development of new analysis tools. Formerly, the discipline relied heavily on mathematical analysis of lower-dimensional, deterministic systems and presented limited computer simulation results. Today, the need to integrate large amounts of variable and uncertain resources from different technologies while maintaining reliability and economic efficiency has dramatically increased the complexity of studying power systems \cite{KROPOSKI2017}. However, thanks to increases in available computational power, optimization-based models and computer simulation are standard tools used for research in systems of all scales, from bulk generation and transmission to micro-grids. This paper contributes a framework to facilitate using these tools in accordance with the principles of Scientific Computing.

Scientific Computing has emerged as a field that studies and promotes the application of principles such as reproducibility, transparency, and accuracy to experiments that are carried out using computer simulations. Although Scientific Computing has benefited from notable contributions regarding the theory and practice of reproducibility \cite{stodden2012reproducible, leveque2012reproducible, bailey2016facilitating, donoho2010invitation}, and there have been crucial advancements in the systematic development of computational experiments for model and algorithm testing through simulations \cite{bartz2010future, kleijnen2005state, sanchez2018work}, its adoption is not widespread.  

It stands to reason that enhancing Scientific Computing would carry significant implications for power systems, since computational experiments afford almost the only option for engineers to conduct empirical experiments about the operation of large-scale power grids. Computing is a fundamental tool for conducting operations research \cite{nestler2011reproducible}, and -- by extension -- power systems operation research. Improving the definitions and practice of Scientific Computing principles thus also stands to benefit both of these fields.

We propose an initial step towards a power systems framework that systematically applies Scientific Computing principles to the development and validation of operational models. {The focus of this paper is operational models, i.e. short-run decision-making models, that commonly represent the system in a quasi-steady state and solved via optimization algorithms. However, the broader concepts and definitions are also applicable to power system dynamic models, electromagnetic transients and any other field that relies on computer simulations.} The relevant contributions in this paper are: first, a template to facilitate the development of simulations and computational experiments. Second, a consistent set of Scientific Computing definitions, practices, and implementation details to facilitate its application to power systems operations research. Third, repositories with cases employing the techniques and definitions outlined in this paper. 

The paper is organized as follows: in Section II, we present a description of Scientific Computing practices and definitions emphasizing their relevance to power systems operations research. The requirements to design a computational experiment focusing on the definition of variables, data, models, and metrics is presented in Section III. Section IV showcases a development pipeline required to implement the computational experiment in accordance with Scientific Computing principles. Section VI includes examples of the implementation guidelines discussed in Section III and IV. Finally, we include conclusions in Sections V.

\section{Scientific computing practices}
\begin{figure*}[t]
	\centering
	\includegraphics[width=2\columnwidth]{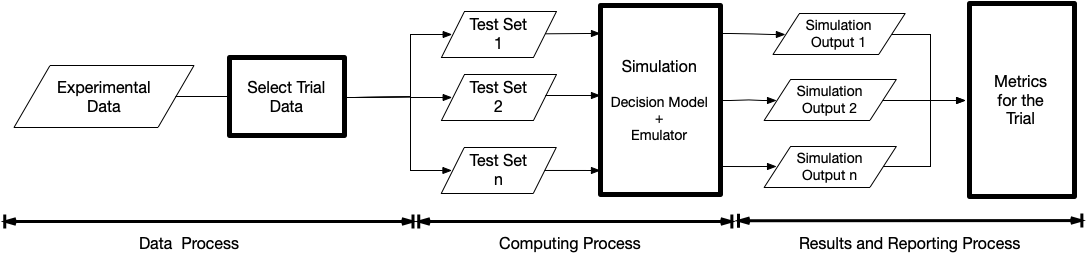}
		\caption{Computing procedure for conducting a single trial}
	\label{fig_trials}
\end{figure*}
The field of Scientific Computing includes a broad array of definitions and practices. The relative importance and specific application of any one of these varies according to discipline \cite{pellizzariETal2017, barba2018terminologies}. Here, we review definitions and applications of Scientific Computing principles in power systems versus other disciplines in the context of the scientific process.

Incorporating new knowledge into established science relies on (1) \textit{reproducibility} of the experiments and (2) the \textit{validity} of conclusions derived. Although there are semantic distinctions across fields in the term reproducibility \cite{pellizzariETal2017, barba2018terminologies}, we define reproducibility as the ability to produce consistent results with repeated experiments while running the same software and using the same input data {and simulation settings}\cite{Goodman341ps12, Peng1226, rougier2017sustainable}. Validity also has a variety of definitions depending on the scientific context. Here, we refer to two types: (1) Internal validity, or the consistency that simulation results have with the experimental system, and (2) external validity, which might also be called ``generalizability." It is the ability to derive inferences about how specific results or relationships may hold over variations in settings and systems \cite{pellizzariETal2017}. These definitions must hold even when the experiments are carried out through simulation in computational environments.  

Interest in Scientific Computing as a field unto itself began with a focus on the precision aspects of computational numerical methods and the conditions required for algorithms to obtain consistent results \cite{heath2018scientific}. But numerical precision is not a chief concern for power systems operation research {as it is for other simulations approaches like dynamic modeling and the development of numerical integration algorithms}. Researchers in this field usually employ third-party packages to solve \acp{MP}, and therefore the responsibility of numerical precision lies with the solver provider. The only exception to this is power systems research focused on the development of algorithms (e.g., \cite{IPM_OPF}).

Some of the most vocal advocates in the scientific community for the principles of reproducibility and validity are from the fields of medicine\cite{Peng1226} and computational biology \cite{donoho2010invitation} which strive to derive consistent, verifiable results from field or laboratory experimental data. Power systems researchers would do well to take lessons applicable to modeling and simulation, since their work is conducted on the basis of numerous embedded assumptions \cite{morrison2018energy}.

Large amounts of resources and time are needed to fully reproduce experiments that are not accompanied by code. However, computational reproducibility requires more than access to the code used to run computational experiments. While documentation alone meets some minimum standard, it is often insufficient to achieve full reproducibility. The recent focus of Scientific Computing on institutionalizing ``best practices" has included themes directly relevant to effectively sharing code and documentation. This includes guidance on code readability, the use of version control tools, and code sharing through platforms like Github. These basic tenets have come to define the baseline for {``good"} code \cite{wilson2014best}. In part, these are not practiced in power systems because most researchers are not trained in the Scientific Computing principles that would emphasize the why and how of reproducible code \cite{benureau2018re}. In many cases, power systems researchers have no intention of re-using code following paper publication. 

In summary, according to the scale of reproducibility proposed by \cite{Peng1226}, most papers in this field thus fall in the category of ``publication only" and can only be regarded as ``reviewable" according to \cite{pellizzariETal2017}. Implementation details are often deemed irrelevant, even when in many cases such details are crucial to understanding final results \cite{stark_before_2018}. By contrast, consider the open source genetics software \texttt{BioConductor} \cite{gentleman2004bioconductor}, whose standard of reproducibility has generated more than 10,000 citations showing the value to the field of having certain common scientific practices. It has no equivalent in power systems, although researchers in this field do occasionally apply principles of Scientific Computing to specific areas, such as the use of \ac{OPF} benchmark cases for the development of new formulations \cite{coffrin2018powermodels}.

\section{Designing the experiment}

In this section, we contribute a logical process for designing power systems computational experiments based on general practices from empirical research. In a simulation context, the experiment consists of varying inputs and executing \textit{trials} of the \textit{simulation processes} to generate a sample of outputs. These samples are later used to compute relevant performance metrics or summary statistics. Figure~\ref{fig_trials} shows a breakdown of the experimental process organized into three different processes: data, computational modeling, and results sub-processes. Each of the sub-processes can present the researchers with many design choices. However,  in this section, we review exemplary practices in the literature to guide tackling common issues and best practices. 
We emphasize that using this framework supports a robust evaluation of operational models and provides a clear way for reviewers and readers to assess conclusions conditional on the researchers' experimental setup.

In \cite{sanchez2018work}, the authors define a simulation experiment as sampling the space of input variables over trials to characterize a system in the space of output variables. Exhaustive sampling {is a viable strategy when the input space is small}. However, power systems operational research typically features complex and large input spaces that includes time-series, network configurations, demand levels, and parameters of economic and physical subsystems. This motivates a sub-classification of the data inputs inspired by general empirical methods to help inform experiment design.

In our review, we find that a formal design approach and classification like this are rarely used. Researchers tend to implicitly focus on reporting a few key simulation results, and over-specify the case data and the experiment parameters, reducing external validity of conclusions.

\begin{table*}[t]
  \centering
\caption{Data Requirements and Best-case Selection Process}
\label{tab:data_criteria}
\begin{tabular}{|l|c|c|c|}
\hline
\textbf{Input Type}                                                       & \textbf{Description}                                                                                             & \textbf{Selection Considerations}                                                                                 & \textbf{Examples}                                                                                                        \\ \hline
\textbf{\begin{tabular}[c]{@{}l@{}}Experiment \\ Parameters\end{tabular}} & \begin{tabular}[c]{@{}c@{}}All data held constant through out the simulations, \\ defining the scope of the experiment\end{tabular} & \begin{tabular}[c]{@{}c@{}} Enables generaliztion \\ of the experiment to other cases\end{tabular}                                        & \begin{tabular}[c]{@{}c@{}}Test network, cost functions,\\  boundaries of the confounding \\ variable space\end{tabular} \\ \hline
\textbf{\begin{tabular}[c]{@{}l@{}}Confounding \\ Variables\end{tabular}} & \begin{tabular}[c]{@{}c@{}}Data varied across trials, used to test \\ robustness of results\end{tabular}         & \begin{tabular}[c]{@{}c@{}}Unbiased sample \\ coverage of the relevant space\end{tabular}                         & \begin{tabular}[c]{@{}c@{}}Forecast and realization \\ time-series, \\ reserve requirements\end{tabular}                 \\ \hline
\textbf{\begin{tabular}[c]{@{}l@{}}Independent \\ Variables\end{tabular}} & \begin{tabular}[c]{@{}c@{}}Variables compared within each trial, \\ primary objects of study\end{tabular}        & \begin{tabular}[c]{@{}c@{}}Isolation of variable \\ of interest, including \\ a control or null case\end{tabular} & \begin{tabular}[c]{@{}c@{}}Operational models, \\ forecast accuracy, \\ renewables penetration\end{tabular}              \\ \hline
\end{tabular}
\end{table*}

\subsection{Data}

The experimental data specification requires selecting system parameters, a method for {selection and use of} trial data, and a definition of \textit{test-sets}. {A test-set is a collection of inputs used in the simulation that produces a corresponding output. The data defining a test-set is organized hierarchically with a single set of \textit{experiment parameters} that define the scope of the experiment, a number of sample sets of \textit{confounding variables} that delineate a particular trial, and for each trial, an instance of each of the \textit{independent variables} which are the primary objects of study with a hypothesized impact on the outputs. Table} \ref{tab:data_criteria} {shows the different types of data input along with descriptions and selection considerations.}

{The validity of results is closely tied to the researchers' choices of both classifying and selecting these data. The more general and representative of other cases the experimental parameters are, the higher the external validity; while the more robust the sampling of confounding variables, the higher the internal validity. The simulation settings such as output interval, algorithm tolerances, convergence criteria are also part of the experiment parameters, and convergence criteria.

Confounding variables can be sampled either randomly or deterministically:} sensitivity analysis, selecting sets of representative or extreme cases, and Monte Carlo simulation are all conventional methods \cite{sanchez2018work}. In power systems operational research, independent variables often {include or are directly the operational models themselves; e.g., the alternative} \ac{UC} models in \cite{wang2011wind,wu2011comparison} or the demand response recourse strategies in \cite{papavasiliou2012stochastic}. Confounding variables are often sets of {time-series data}, initial system state, renewable penetration or the test system.

{In selecting data,} there is often a tension between using real, benchmark, or synthetic data. Real systems and real data sets are often attractive because they have better internal validity. However, using any proprietary data with sharing restrictions can undermine reproducibility. Standardized test systems are by definition more reproducible and should yield more comparable results across studies, but the prevalence of ``modified IEEE systems'' show that these often do not capture all relevant features and there is a need for comprehensive test data sets. Synthetic network generation algorithms address some of these shortcomings \cite{BirchfieldSyntch}, but do not eliminate the need for modification with time series and the addition of other devices. In these cases, the modification process must be transparent and unbiased, which can be done by {providing all the details} when defining the modifications necessary and generating the full modification with code. Capturing uncertainty in synthetic data sets, particularly in forecasts and realizations is a critical component in modern operations; \cite{quan2015computational} provides an overview of modeling uncertainty across parts of the power system. Regardless of the data source or generation method, access to and interpretation of the data used is necessary to comply with standards of reproducibility and transparency. To our knowledge, only the recent update to the RTS-96 data set \cite{BarrowsRTS} complies with these requirements. {Further, scientific principles must be followed by not ``cherry-picking'' data to produce favorable results. Adherence to this framework and detailing data selection processes can make it harder to obscure this behavior, but integrity is still required of the researcher.} 

\subsection{Simulation Models}

In power systems research, computational models are both the subject of the research and the experimental apparatus for evaluating research, often leading to the conflation of different models within the experiment.
In practice, most experiments are used to demonstrate the performance of a proposed \textit{decision model} for taking an operational action. In our framework, this makes the decision model an independent variable. Following the process in Fig. \ref{fig_trials}, it is necessary to define at least one alternative test-set (i.e., a control) in the form of a baseline decision model, often representing a heuristic or current practice. To do this, and to have a valid representation, it is critical to delineate a \textit{decision model} from an \textit{emulator model} used as a representation of the real system defined as follows:

\begin{itemize}
    \item {Decision Model: The model used to obtain the desired system operation behaviour. The model that generates set-points or policies used to drive the devices in the system.}   
    \item {Emulator Model: A model that mimics a specific real-world behaviour of the electric power system. The model produces outputs that resemble the system performance when operating under the resulting policies from the decision models.}
\end{itemize}

The emulator yields the performance metrics to compare performance between decision models and should be consistent across all test-sets and trials. 

{In some cases, where the contributions are comprised of a combination of operational models and solution algorithms} (e.g. \cite{nasri2015network, zhang2017chance}) {it is important to have a clear distinction between the modeling aspects and the algorithmic ones. In such cases, details of the model and implementation of the algorithm should be clear and reproducible. For instance, update rates, constants, tolerances and heuristics need to be clearly stated.}

The results commonly included in publication consist of the output of the decision model without showing whether the decision model has the hypothesized effect on the system. The  emulator model is intended to enable the evaluation of the decision model's performance in terms of the system's behavior conditional on its output.  Not having an emulator in the simulation process is particularly problematic in operations models intended to handle uncertainty like \ac{SUC} and \ac{RUC}. It also complicates matters when the decision model must make simplifying assumptions to be computationally tractable. In \cite{tuohy2009unit}, the authors explicitly point out the limitations of interpreting the results of their decision model, and propose a modification to make it more realistic. However, the standard should not be that a decision model accurately represents reality; rather, the effort is selecting a reasonably accurate emulator to capture the relevant system's performance.

Separating the models in this fashion has been done in many highly cited works \cite{papavasiliou2012stochastic,wang2011wind,tuohy2009unit}, though there are some exceptions \cite{wu2011comparison}. Other examples are most readily found in receding and shrinking horizon model predictive control, often employed in microgrid \ac{EMS}. The stochastic \ac{EMS} work in \cite{parisio2014model} uses an actual physical test system, while \cite{su2013stochastic} exemplifies the computational approach using OpenDSS as the emulator, and \cite{lara2018robust} uses a \ac{MC} approach to test the robustness of the control.

Based on the aforementioned literature, we generalize three main characteristics that the emulator should possess to be used as the validation test-bed in simulations: 

\begin{enumerate}
    \item The emulator should be on the same or a faster time-scale than the decision model under study, and capture the phenomena that are significant on these time scales and relevant to the study. There needs to be a logical connection between the chosen emulator and the operational model.
    \item The system representation should include as much detail as is necessary to test the effect of simplifying assumptions used to make operational models tractable. This often requires additional data and trials.
    \item The emulator's time-series realization data \textit{must} be distinct from forecasts used in the decision model. The use of forecasts within decision models has become commonplace in power systems operations research, and in many cases the same data used to generate the decision are used to test the effectiveness of the decision model.
\end{enumerate}

\subsection{Performance Metrics}

Metrics are measurements computed on outputs of the computing process in Fig \ref{fig_trials}. The relevant aspect to consider is that metrics should be reported and calculated in terms of the trials with a probabilistic sense. Commonly, metrics are reported only for a single test set or single trial. In a computational experiment conducted with repeated trials, the metrics should accurately demonstrate the internal and external validity of the conclusions. For instance, in \cite{wang2016quantifying, martinez2016value} the authors use total costs and Automatic Generation Control performance metrics (e.g., CPS2) to assess the value of improving forecasting accuracy in the system.

There are a vast number of possible metrics relevant to power systems operations that depend on the application of the decision model. However, it is possible to derive recommend three principles in the design of experimental metrics:
\begin{enumerate}
	\item Consider the distribution of metrics across trials and analyze them statistically.
	\item Disaggregate metrics rather than reporting total costs alone. Presenting the costs associated with here-and-now and recourse actions separately can yield more insight (neither can be omitted), and include physical metrics which provide a richer picture of phenomena left out of decision models but captured by the emulator (e.g., battery wear, insecure loading limits, etc.).
	\item Include the computational times for the given environment. Metrics on computational performance tend to go unreported and they are critical for assessing the feasibility of incorporating research into real systems or further experiments, as well as indicating opportunities for computational performance improvement.
\end{enumerate}

\section{Implementing a reproducible computational experiment in power systems}

In practical terms, achieving reproducibility and validation requires operationalizing the principles of Scientific Computing. Here, we present a template for a reproducible scientific workflow {for simulation of power systems operations problems}. Each step of this template weighs the two major components required for a computational experiment: the \textit{environment} and the \textit{workflow}.

\begin{itemize}
    \item \textit{Environment:} The collection of software, hardware components, and configurations used to implement a computational experiment \cite{benureau2018re}. The environment may include elements such as cloud-services, third-party software, file management scripts, and external tests.
    \item \textit{Workflow:} The sequence of computing tasks, from data intake to summary of results via plot and table generation, which, as a collection, make up the scientific experiment and the analysis \cite{deelman2018future}.  The development of a workflow is a requirement for validation and reproducibility. 
\end{itemize}

The following template breaks down both the environment and the workflow of a power systems computational experiment into the three main stages: \textit{data process}, \textit{computing process}, and \textit{results and reporting process}. 

\subsection{Data process}

Often, preoccupations about data in the context of Scientific Computing are limited to data sharing, access concerns, and best practices for transparency by using raw data files in human-readable formats \cite{wilson2014best}, such as CSV for tabular data. However, the very act of processing raw data into analytical data must itself be a reproducible procedure that can be validated. The two aspects necessary for achieving this are (1) the data model and (2) the data production process. 

\begin{figure}[t]
	\centering
	\includegraphics[width=0.75\columnwidth]{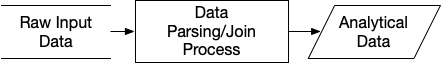}
	\caption{Data Processing.}
	\label{data_flow}
\end{figure}

All computational experiments need to define, document, and automate the steps followed in processing and generating data, such that these procedures can be reproduced and validated by other researchers. Figure \ref{data_flow} showcases a simple workflow for data processing and generation, distinguishing between the \textit{raw input data} that can be obtained from many different sources and may exist in a variety of formats, and the \textit{analytical data} that has been parsed by the researcher to fit neatly into the organization of the data model. 

\subsubsection{Data Model}

A data model is a way to organize data and standardize its internal relationships and properties{, providing a structure for use in the computational experiment} \cite{west2011developing}. In power systems research, efforts to make standard data models have primarily focused on power flow data \cite{cf}, and the Common Information Model (CIM) developed by industry for SCADA and control center automation \cite{uslar2012common}. Power systems' operational research requires richer data models, which can hold more information than just a representation of the system parameters. For instance, holding time series, confounding variables, and parameters is necessary for the execution of computational experiments.  

Without a single, widely agreed-upon data model in power systems modeling for computational experimentation, researchers have to develop customized data models for individual applications. {This process is time-consuming, and} as a result, data {models} usually go underdeveloped. {Researchers should first carefully evaluate whether a custom data model is necessary. If so,} modern programming languages provide researchers with environments containing an extensive assortment of options to develop data models. {Researchers can greatly increase the value of their contribution and convergence to new standards by publishing their data model and implementation code for re-use.}

\subsubsection{Data {C}onsumption}

In most cases, before the data can be arranged into the data model, some computations are necessary. Examples of such computations might include consistency checks, ensuring the anonymity of proprietary data, or adding more features to the original data. For instance, including ramp rates to model \ac{UC} in data sets that were originally developed for \ac{PF}. Such computations are part of the data processing pipeline, and must be recorded as part of the workflow. 

In the context of the workflow depicted in Figure \ref{data_flow}, synthetic data should be considered, along with ``real" or observational data, to be a raw input to the model. Like real data, synthetic data may undergo parsing for use with a specific data model. When creating confounding data, all random number generators used should be seeded, and those seeds included in the data model. Further details concerning the process for generating synthetic data are outlined in Section~III.

\subsection{Computing {P}rocess}

\begin{figure}[t]
	\centering
	\includegraphics[width=0.95\columnwidth]{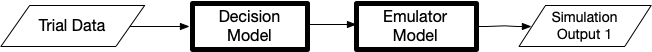}
	\caption{Simulation execution workflow.}
	\label{sim_flow}
\end{figure}

\begin{figure}[t!]
	\centering
	\includegraphics[width=0.95\columnwidth]{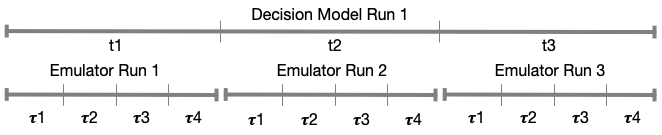}
	\caption{Synchronized simulation model.}
	\label{synch_tl}
\end{figure}

\begin{figure}[t!]
	\centering
	\includegraphics[width=0.95\columnwidth]{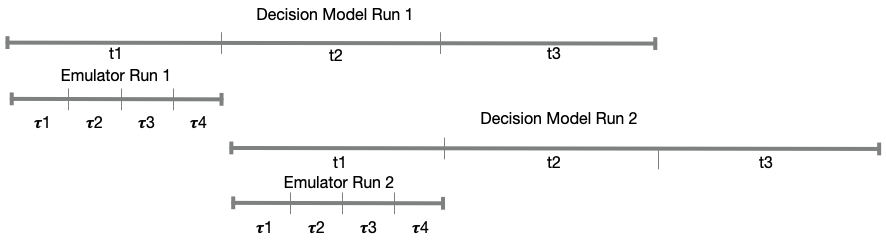}
	\caption{Receding Horizon simulation model.}
	\label{rh_tl}
\end{figure}

The computing process is easily confused with the computational experiment itself. In fact, the experiment -- as described earlier in this paper -- is far more nuanced than the execution of a model. A computing process is comprised of a set of simulations used to test the effects on the test-set in the system using the emulator model as a proxy, as shown in Fig. \ref{sim_flow}. Each trial of an experiment constitutes a simulation case, which is characterized by the decision and emulator model executions and interactions. For instance, simulating a standard day-ahead \ac{UC} over a year results in executing 365 daily decision models, each followed by another 24 hourly \ac{ED} emulator executions.  

Simulations can take many different configurations depending on the research objective and the experiment design. It is, therefore, not possible to pre-define a generic simulation setup that fits all applications. However, it is possible to establish some general definitions to help guide the development and facilitate the reproducibility of the simulation setup. 

A simulation generally defines a chronology in which models execute, and information is passed between them. Although there are myriad configurations in power systems, there are two standard formats: 1) \textit{Synchronized}, where the results of each time-step \textit{t} in model A is synchronized with the execution of model B such as the day-ahead example mentioned above (see Fig. \ref{synch_tl}), and 2) \textit{Receding Horizon} where only the solution of $t =1$ is used in model B such as Example 1 in Section V (see Fig. \ref{rh_tl}).  

\subsubsection{Model Execution}

\begin{figure}[t]
\centering
\includegraphics[width=0.95\columnwidth]{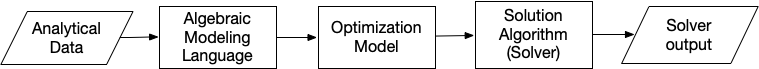}
\caption{Optimization Model Execution Pipeline.}
\label{optimization_pipeline}
\end{figure}

When an optimization model is executed, sub-processes relevant pieces to scientific workflow happen in the background that need to be accounted for in the computation workflow. Figure \ref{optimization_pipeline} shows the pipeline and the different components that make up the environment and workflow of a model execution.

\begin{itemize}
	\item \textit{\acl{AML}}: The \ac{AML} code describes the model's \ac{MP}. The models are generally expressed in terms of \textit{sets}, \textit{parameters}, \textit{variables}, \textit{constraints}, and an \textit{objective function}. In this stage the optimization model is not being solved, rather its processed such that it can be used in the solver.
	
\item \textit{Optimization Model}: This is the \ac{AML} output, the optimization model in standard form. Often \acp{AML} make internal transformations to the \ac{MP} to make it compatible with the solver. This include adding slack variables or constraint reformulations, and can occasionally hinder reproducibility when the \ac{AML} changes.

\item \textit{Solver}: The solver performs the computations necessary to obtain the solution of the \ac{MP}. Modern optimization algorithms can be heavily customized, and all the details of the solver parametrization must be accounted for.
\end{itemize}

\subsection{Results and Reporting}

\begin{figure}[t!]
	\centering
	\includegraphics[width=0.85\columnwidth]{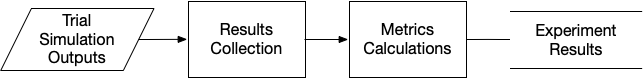}
	\caption{Results and Reporting Workflow.}
	\label{fig_results_reports}
\end{figure}

The output of the computing process is the collected results of the simulations, which does not necessarily imply that it is the final result of the experiment. Figure \ref{fig_results_reports} shows the workflow to compile the outputs of the model executions into a format suitable for analysis, finishing with the calculation of the selected metrics. Often, other libraries are required in the environment to process the outputs and calculate the results. The aggregation process has the potential to confound the effects of the proposed model. Hence, the raw output of the simulations should be archived, and the code to calculate the metrics should comply with the principles of reproduciblity.  

\section{Case studies of the framework implementation}
We have included two example cases for the principles described in this paper. {The cases are developed in Matlab and Julia programming languages to showcase that the proposed framework is platform and language independent.} The results shown here aim to exemplify the application of the principles in simplified case studies and not to derive specific conclusions about the models. The code for the case studies is archived in the public repository
 \url{https://github.com/Energy-MAC/PowerSystemsScientificComputing}.  

\subsection{Case 1: Microgrid EMS with demand response}
This experiment compares different predictive controllers to regulate power consumption through demand response in energy-constrained islanded microgrids.

\subsubsection{Experiment Design}
The hypothesis is that each controller improves a particular performance metric relative to a baseline of no control. Here, the choice of decision model used by the controller is the independent variable. We compared four controllers: a heuristic method that does not use forecasts, a predictive control using only a single average forecast, and two stochastic programming formulations. 

We constructed a synthetic microgrid, synthetic forecasts of solar generation based on historical data, and synthetic electricity demand forecasts based on random simulation to be used as confounding variables. The system features multiple customers, distributed solar and battery storage.

The performance metrics are:  Average Service Availability Index (ASAI), total customer utility from electricity consumption, and the realization of the objective of the decision model, which is a measure of benefit of electricity consumption that we construct. Since performance metrics include power availability and the impact of interruptions at the customer level, the emulator must capture the effect on customers.

\subsubsection{Implementation}
This environment is MATLAB 2018a and Gurobi 9.01 as the solver using the Gurobi MATLAB API. The experiment uses a custom data model for data input/output and for the experimental parameters. 
The controller makes a decision to send energy and power limit signals to customers every 4 hours with a receding horizon of 2 days. The emulator model includes a customer decision to respond to the signal every 4 hours, and a model of the physical devices that runs on a 2-minute timescale.
The controller assumes that customers will reduce consumption if necessary to comply with energy and power limits, but assumes they will reduce \textit{exactly} to a limit, while in the emulator, customers make a discrete choice of which appliances to disconnect with some imperfect knowledge of what their consumption will be.
Then, the 2-minute model simulates power sharing between generation and storage, the evolution of thermostatically controlled loads and battery state-of-charge, and interruptions when load power cannot be met or customers exceed limits.
\subsubsection{Results}
\begin{figure}[t]
\centering
\includegraphics[width=\columnwidth]{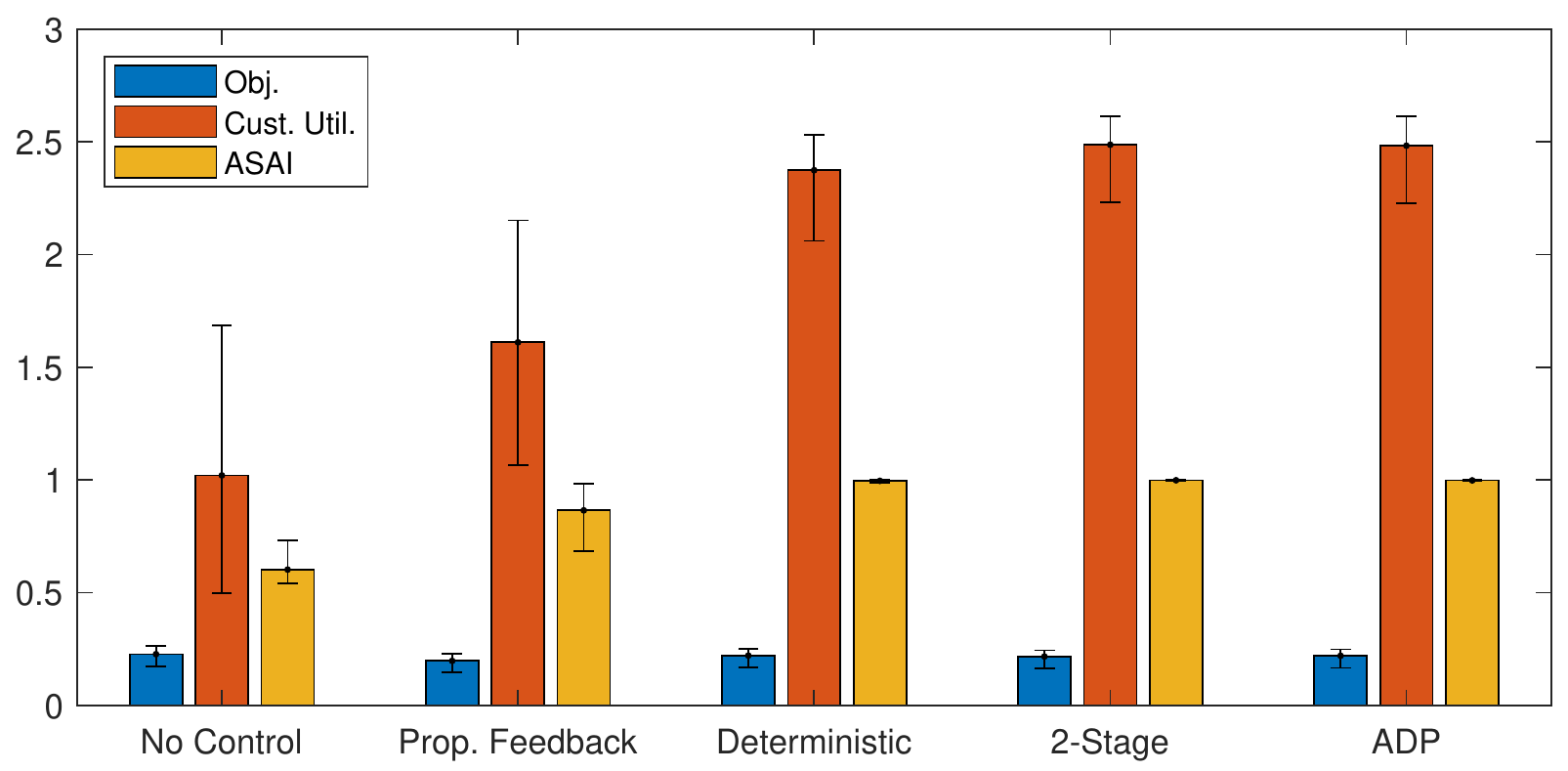}
\caption{Performance metrics for different controllers for Experiment 1 as median values with 5th and 95th percentile error bars}
\label{fig:performance_exp1}
\end{figure}

Figure \ref{fig:performance_exp1} shows the value of the metrics across trials for each controller relative to the baseline of no control. This illustrates the principles of displaying metrics statistically and showing the side effects on multiple concrete emulator metrics not explicitly optimized for in the decision model that has an abstract objective. The design framework emphasizes that these results are limited to the context of the experiment parameters, particularly the interruption cost function in the emulator customer model and the forecast and realization data.
%In fact, following the principles outlined in this paper revealed initial poor performance with using power limits that we were able to address by instead using an energy limit, but would have been obscured without a higher time-resolution emulator model.

\subsection{Case 2: Stochastic UC and Bulk Power Systems}
 The experiment is designed to compare an \ac{SUC} model with a standard \ac{UC} model. The detailed mathematical specifications of the models used in these experiments are located in the example repository.  

\subsubsection{Experiment Design}
The hypothesis is that using \ac{SUC} achieves better hedging against wind power uncertainty and reduces the occurrence of load shedding.
In this example, the independent variable is the day-ahead model which is the choice of \ac{UC} or \ac{SUC} model. Given that the \ac{SUC} requires the use of scenarios, the complete test-set is made up of the decision model-forecast pair. The \ac{SUC} uses 100 scenarios generated by sampling a truncated normal distribution with a variance of 30\%. To evaluate the hypothesis, an \ac{ED} model is used as the emulator with 5-minute resolution time-series. 

The experiment includes 10 trials each for ranges of 30 consecutive days in the annual data in order to capture seasonal variations through the year. For each trial, the metrics used to assess the performance of the system are total fuel cost and total energy not supplied. The models are tested in a modified 5-bus test system \cite{Li5Bus} which has been enhanced with piece-wise linear cost functions. The test system also features an additional 3,600 MW of wind power generation and an increased peak load of 14,400 MW. The system has been also enhanced with yearly time series for load and wind power from the data provided in \cite{BarrowsRTS}. 

\subsubsection{Implementation}

The experiment has been in Julia v1.2.0, using the JuMP v0.20.0 as the \ac{AML} with the Solver Gurobi 8.11. Auxiliary \texttt{*.toml} files define the full experiment environment by fixing other libraries' versions in the experiment. The code structure in the example repository has been intentionally arranged consistently with the definitions given in the paper. The data model is provided using the package \texttt{PowerSystems.jl} and the same is used to integrate time series data into the load flow case. The commitment decisions are synchronized with the \ac{ED} model, which is executed 24 times for every commitment model execution.  

\begin{figure}[t]
\centering
\includegraphics[width=\columnwidth]{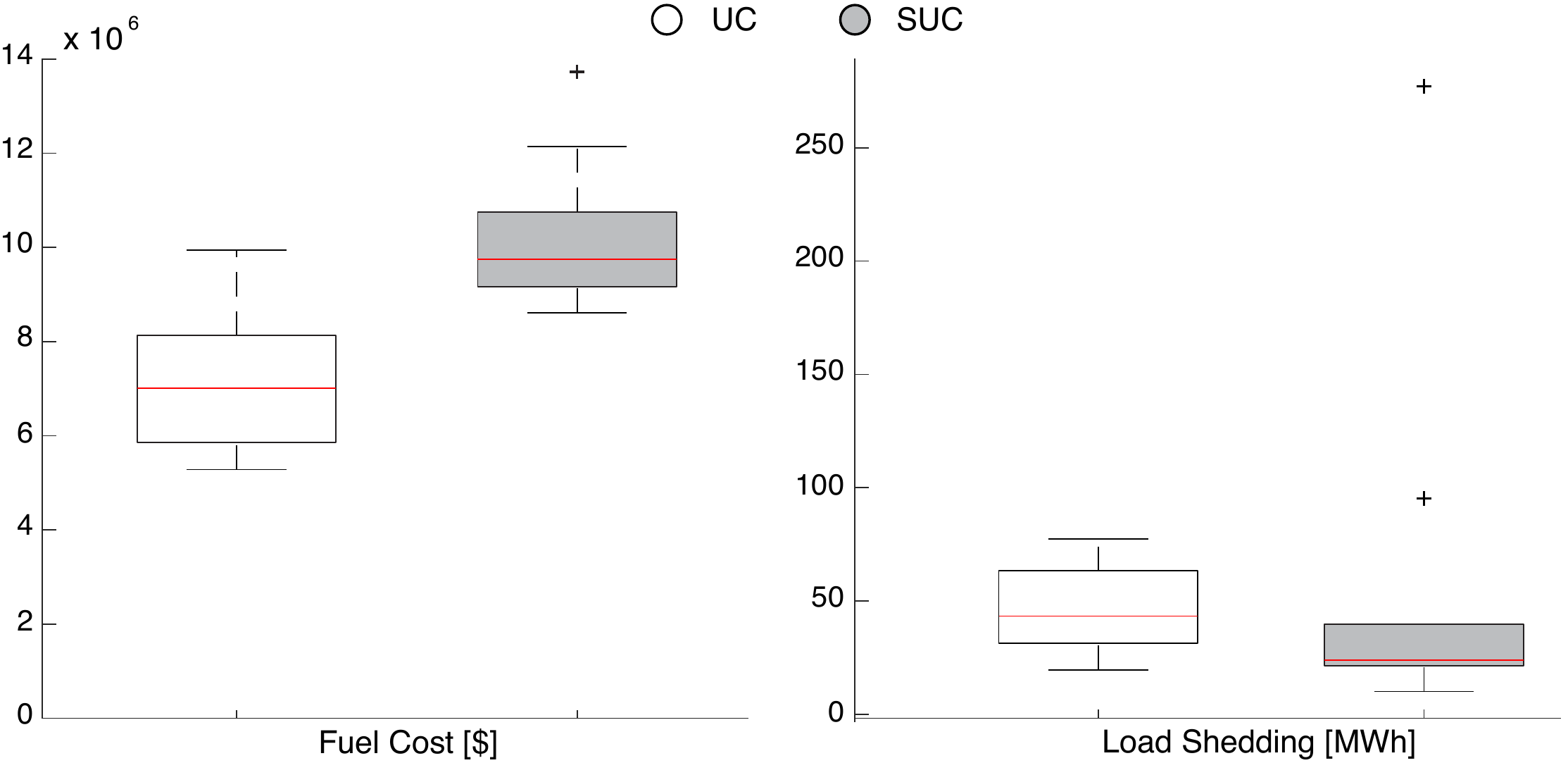}
\caption{Box plots of results for Experiment 2}
\label{fig:boxplots}
\end{figure}

\subsubsection{Results}

The results showed that for most trials, the \ac{SUC} reduces the amount of load shedding in the system but also has an increased fuel cost. The results are displayed as Box Plots {in Figure} \ref{fig:boxplots} that enable further exploration of the trial results and derive more detailed insights about the model's performance. In this way, it is possible to notice that in two outlying samples, the \ac{SUC} resulted in more load shedding than the \ac{UC}.  The result motivates exploring which circumstances may lead to weaker performance from the \ac{SUC} than the \ac{UC}. 

From the computational point of view, running multiple trials allows us to evaluate if the relative differences in computation speed are consistent between the model. Each commitment model ran a total of 300 times; the \ac{SUC} has an average solve time of 59.18 s and a maximum of 158.98 s compared to the 2.87 s and a maximum of 8.2 s of the \ac{UC} model; an order of magnitude faster than the \ac{SUC}.

\section{Conclusions}

\begin{itemize}
    \item The power systems community has the potential for many benefits from the adoption of Scientific Computing practices. The current computational experimentation practices are lacking when compared to other fields.
	\item Developing novel operational models requires validation through computational experiments. Appropriate experiment design is critical in demonstrating models' capability to handle large penetration of renewable energy.  
	\item The software developed for power systems operational research must reproduce experimental results, which implies access to the input data, code access and consistent results when the experiment is replicated.
\end{itemize}


%\bibliographystyle{IEEEtran}
\bibliographystyle{IEEEtran}
\bibliography{PowerSystemsScientificComputing}
%

\end{document}